\documentstyle[11pt,twoside,newpasp,epsf]{article}

\begin{document}
\title{Photorecombination and Photoionization Experiments at Heavy-Ion Storage-Rings and Synchrotron-Light Sources}
 \author{Stefan Schippers}
\affil{Institut f\"{u}r Kernphysik, Justus-Liebig-Universit\"{a}t Giessen, Leihgesterner Weg 217, 35392 Giessen, Germany}

\begin{abstract}
Recent experimental work on the photorecombination and the photoionization of
astrophysically relevant atomic ions employing the merged-beams technique at
heavy-ion storage-rings and synchrotron-light sources, respectively, is summarized.
The resulting {\em absolute} photoionization cross sections and recombination rate
coefficients benchmark corresponding theoretical calculations and are needed for the
accurate determination of ionization equilibria in astrophysical plasmas.
\end{abstract}

\section{Introduction}

Photoionization (PI) and Photorecombination (PR) are basic atomic processes which govern the
charge state balance in any plasma. Therefore, the accurate knowledge of these processes  is a
prerequisite for any plasma modeling and, hence, for any meaningful interpretation of many
astrophysical observations. To date most PI and PR cross sections used in plasma modeling stem
from theoretical calculations. Their accuracy is often difficult to asses and, therefore,
experimental PI and PR cross sections and rate coefficients are vitally needed as benchmarks
and guidelines for the development of the theoretical methods. Moreover, in the near future
experiments will probably be the only reliable source for PR and PI cross sections of complex
ions such as ions with an open M-shell (Schippers et al.\ 2002a; 2003a).

The measurement of PR and PI cross sections of atomic (and molecular) ions is experimentally
challenging because the ion densities which can experimentally be prepared are very low ---
some 10$^6$~cm$^{-3}$ (to be compared with $\sim 10^{13}$~cm$^{-3}$ for gaseous and $\sim
10^{22}$~cm$^{-3}$ for solid targets). Consequently, signal rates from electron-ion and
photon-ion collision experiments are comparatively weak. In order to make up for the low ion
density in such experiments an arrangement is chosen where the colliding particle beams are
merged collinearly over a distance of the order of 1~m. The merged-beams arrangement (Phaneuf
et al.\ 1999) provides a relatively large interaction volume, and the directionality of the
ion beam facilitates the effective collection of the PI or PR reaction products, i.\,e., of
ions which have changed their charge state due to either ionization or recombination. For the
measurement of {\it absolute} PR and PI cross sections the merged-beams method was implemented
at heavy-ion storage rings (M\"{u}ller \& Wolf 1997; M\"{u}ller \& Schippers 2001) and at
synchrotron-light sources (West 2001; Covington et al.\ 2002), respectively.

\section{Photorecombination of atomic ions}

\begin{figure}[hhh]
 \plotfiddle{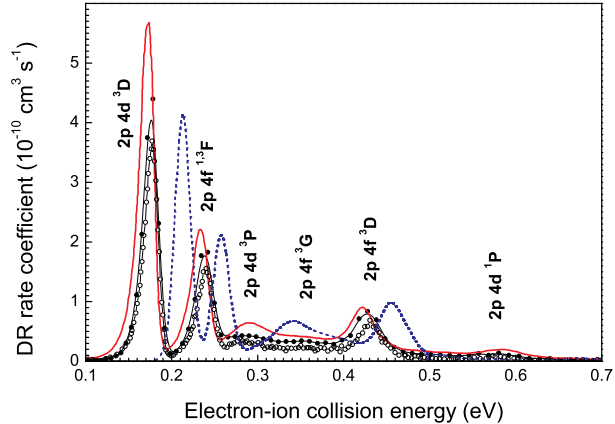}{4.8cm}{0}{70}{70}{-140}{-20}
\caption{Dielectronic recombination (DR) of C$^{3+}$($1s^2\,2s$)-ions at low energies:
experiment (full symbols: Schippers et al.\ 2001, open symbols: Mannervik et al.\ 1998), many
body perturbation theory (full line: Mannervik et al.\ 1998) and Breit-Pauli R-Matrix theory
(dashed line: Pradhan et al.\ 2001). }
\end{figure}

The basic procedure for producing plasma rate coefficients from PR measurements at storage
rings was outlined by M\"{u}ller (1999). Recent examples are the PR rate coefficients of the
lithiumlike ions C\,{\sc iv} (Schippers et al.\ 2001), O\,{\sc vi} (B\"{o}hm et al.\ 2003) and
Ni\,{\sc xxvi} (Schippers et al.\ 2000). Figure 1 shows a comparison between measured and
calculated C\,{\sc iv} dielectronic recombination (DR) rate coefficients at low electron-ion
collision energies. The two experimental results, which were obtained a two different
storage-rings, agree with one another to within the 15\% experimental error for the absolute
cross section. This exemplifies the reliability of recombination measurements at storage
rings. Mannervik et al.\ (1998) also calculated the C\,{\sc iv} DR rate coefficient using
relativistic many-body perturbation theory (RMBPT, Tokman et al.\ 2002). They pointed out that
even for a light ion such as C\,{\sc iv}, a {\it relativistic} treatment of the recombination
process is necessary to reproduce the experimentally observed $2p\,4l$ DR resonance
structure. Presently, RMBPT is only applicable to relatively simple systems and it is
therefore instructive to see in how far more standard "production" codes are able to reproduce
low-energy  resonance positions and strength. For C\,{\sc iv} the Breit-Pauli R-Matrix results
of Pradhan et al.\ (2001) are only up to 20 meV off the measured resonance positions (figure
1). It should be kept in mind, however, that there are pathological systems, e.\,g.\ Mg\,{\sc
ix}, where a mere 50~meV uncertainty of low-energy resonance positions can translate into a
nearly one-order-of-magnitude uncertainty on the plasma rate-coefficient scale in a
temperature range where the ion is expected to form in photoionized gases (Schippers et al.\
2004). Such findings strongly emphasize the need for experimental benchmarks, especially for
low-temperature PR rate coefficients.

Photorecombination measurements responding to astrophysical data needs were carried out for
L-shell iron ions at the Heidelberg storage-ring TSR by Savin et al. (1997; 1999; 2002a;
2002b; 2003) who already published DR rate coefficients for the ions Fe\,{\sc xviii} through
Fe\,{\sc xxii}. In a series of measurements with lithiumlike ions the influence of external
electromagnetic fields on DR cross sections was thoroughly investigated by Bartsch et al.\
(1997; 1999; 2000), B\"{o}hm et al.\ (2001; 2002) and Schippers et al.\ (2000). This work
is also of astrophysical interest. It was recently summarized by M\"{u}ller \& Schippers (2003).

\section{Photoionization of atomic ions}

\begin{figure}[hhh]
 \plotfiddle{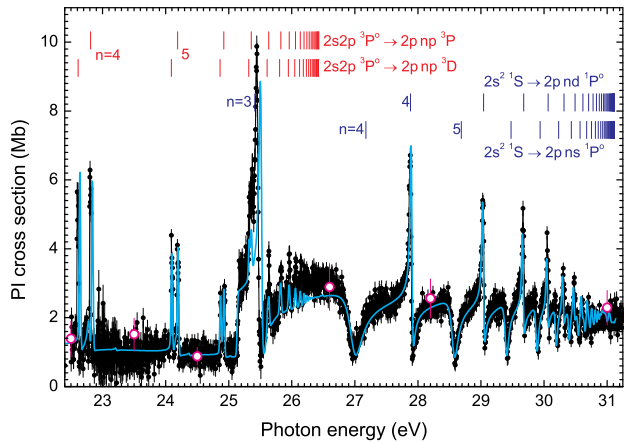}{4.8cm}{0}{70}{70}{-140}{-20}
 \caption{Photoionization of B$^+$-ions: experiment and Breit-Pauli R-Matrix theory
 (Schippers et al.\ 2003b). From the
 comparison between experiment and theory it is concluded that 29\% of the ions in the experiment were
 in the metastable state.}
\end{figure}

More recently experiments aiming at measuring absolute PI cross sections of atomic ions and
employing merged photon-ion beams were set up at 3rd generation synchrotron light sources.
Measurements of astrophysical relevance were carried out e.\,g.\ for singly charged boron
(figure 2, Schippers et al.\ 2003b), carbon (Kjeldsen et al.\ 2001), nitrogen (Kjeldsen et
al.\ 2002a), oxygen (Covington et al.\ 2001; Kjeldsen et al.\ 2002a; Aguilar et al.\ 2003a),
neon (Covington et al.\ 2002), magnesium (Kjeldsen et al. 2000; Aguilar et al.\ 2003b) and
iron (Kjeldsen et al.\ 2002b) as well as for the multiply charged ions C\,{\sc iii} (M\"{u}ller et
al.\ 2002), Ne\,{\sc iv} (Aguilar 2003c), and Al\,{\sc iii} (Aguilar et al.\ 2003b).

Finally, it should be mentioned that the possibility to relate the time-inverse processes of
PI and PR via the principle of detailed balance can be exploited for consistency checks
between PI and PR measurements (M\"{u}ller et al.\ 2002) and to obtain more comprehensive results
than from only one experiment alone (Schippers et al.\ 2002b, 2003a).

\acknowledgements

The experimental work summarized in this article was carried out in close collaborations with
S. B\"{o}hm, C. Brandau, S. Kieslich, W. Shi, and A. M\"{u}ller (Gie{\ss}en), D. W. Savin (New York), G.
Gwinner, M. Schnell, D. Schwalm, and A. Wolf (Germany), H. Danared, N. Ekl\"{o}w, and R. Schuch
(Stockholm), A. Aguilar, A. M. Covington, E. D. Emmons, M. F. Gharaibeh, and R. A. Phaneuf
(Reno).

\end{document}